# Post Quantum Secure Command and Control of Mobile Agents

Inserting quantum-resistant encryption schemes in the Secure Robot Operating System

Richa Varma, Chris Melville, Claudio Pinello, Tuhin Sahai
Raytheon Technologies Research Center
Berkeley California, USA
richa.varma@rtx.com, chris.melville@rtx.com, claudio.pinello@rtx.com, tuhin.sahai@rtx.com

*Abstract*—The secure command & control (C&C) of mobile agents arises in various settings including unmanned aerial vehicles, single pilot operations in commercial settings, and mobile robots to name a few. As more and more of these applications get integrated into aerospace and defense use cases, the security of the communication channel between the ground station and the mobile agent is of increasing importance. The development of quantum computing devices poses a unique threat to secure communications due to the vulnerability of asymmetric ciphers to Shor's algorithm. Given the active development of new quantum resistant encryption techniques, we report the *first* integration of post-quantum secure encryption schemes with robotic operating system (ROS) and C&C of mobile agents, in general. We integrate these schemes in the application and network layers, and study the performance of these methods by comparing them to present-day security schemes such as the widely used RSA algorithm.

## I. Introduction

This paper deals with three major trends in the cyber-physical systems domain: the *increased attack surface* of these systems due to the explosive growth of Internet-enabled devices in the last two decades; the *establishment of the Robotic Operating System (ROS)* as a powerful enabler of robotic and AI systems research, and more and more commonly also in industrial applications as a quick and reliable way to jumpstart a project/product; and the *progress in quantum computing* technologies which poses a potential risk to the security of several existing encryption, authentication, and key exchange schemes, notably the ubiquitous RSA [1] and similar public-key crypto systems. Some sources [2] estimate that the quantum computing threat to traditional public-key cryptosystem will become significant by 2040, which implies that data and credentials need to start being protected a lot sooner to prevent store-and-attack schemes. This looming quantum threat to traditional crypto systems will impact all aspects of communication, storage, credentialing, and the entire lifecycle of commercial, industrial, aerospace and defense systems.

As described later, standard versions of ROS do not provide security for its middleware layer, and any ROS instance with access to the network where the ROS master resides, can also access communication channels including topics and services used by other nodes and ROS instances. Although there is a growing body of work on adding security to ROS, given previous considerations, it is imperative that these solutions address post-quantum vulnerabilities.

Our manuscript is organized as follows. In Section II.A we discuss previous work in the area of security enhancements to the application and network layers of ROS systems. In Section II.B we provide an overview of promising state-of-the art post quantum security schemes and their advantages over the more widely used authentication and encryption methods in today's world. In Section III, we discuss our work on integrating existing post-quantum resistant cryptosystems at the ROS application layer. This enables designers to quickly retrofit existing ROS systems to protect a subset of the critical communications from attacks by untrusted ROS computers, without having to migrate all nodes to a newer/network-secure ROS version. In Section IV, we illustrate our work on fielding a post-quantum secure scheme by leveraging Secure ROS[1] and an implementation of a post quantum-secure crypto system. *To the best of our knowledge, these demonstrations using the application and network layers are the first published instance of post-quantum secure command & control of mobile agents.* The relevance to modern and future aerospace and industrial systems is clear (e.g. unmanned aircrafts, and safety critical industrial processes). Note that our use of ROS is motivated by its popularity and the demonstration of post quantum C&C of mobile agents can be ported to other applications and settings.

In Section III.C we present experimental data of running a PQS crypto authentication system on an embedded platform and show a performance comparison between the PQS authentication and the widely used RSA schemes. While some PQS schemes can be more computationally demanding than traditional systems for comparably-sized keys, they do achieve PQS at practical key sizes, while traditional systems could never be PQS at any key size, thus rendering any complexity advantage inapplicable. In Section IV.C we compare existing

This work was supported by Raytheon Technologies Research Center Internal R&D (IRAD) funds.
[1]http://secure-ros.csl.sri.com/

IPSec based VPN with post-quantum secure counterparts within ROS, and describe its advantages for authorization enforcement. Finally, in Section V, we give an overview of some future research directions.

## II. RELATED WORK

### A. Security enhancements for Robot Operating System

ROS has been widely used in the robotics community for research and development of a variety of robotic and AI systems. ROS was originally aimed at the research community and consequently, little or no thought was given to cybersecurity considerations. The idea was that ROS-based systems would only operate in laboratory-like environments and isolated networks that are not connected to the Internet. However, over the years, ROS has become increasingly popular in the industry with applications ranging from industrial manipulators, sensors and device networks to service robots [3]. Thus, it becomes necessary to incorporate security mechanisms either within ROS or associated system software.

One solution is to add security in the application layer [4] [5] [6]. This work uses a handshake protocol and an authentication server along with dedicated ROS node functions to implement security measures. These methods, however, exploit RSA and AES for authentication. Our work takes the *critical step* of bringing post-quantum safe authentication algorithms to the application and network layers of the ROS architecture.

Other works introduce security mechanisms within the core ROS packages, such that the users of ROS do not have to be concerned with the implementation of such mechanisms. The Open Source Robotics Foundation (OSRF) has developed the SROS project [7] that uses TLS to secure TCP channels. Access control for topics, services and parameters is provided using policies. SROS uses a key-server to generate and distribute certificates during an initialization phases. Although it has been in development for several years, it has not yet been incorporated as a default in ROS and is still highly experimental, with sparse documentation.

Secure ROS, developed by SRI is an approach to secure the ROS core on the network layer, in addition to modifying core ROS packages for access control. IPSec is used to secure the communication channels at the Internet layer. A YAML file is used to prescribe access for nodes, services, and parameters. Secure ROS is relatively easy to incorporate in existing ROS systems and provides clear documentation for reuse. We, therefore, build upon Secure ROS and modify it to use post-quantum secure communication between multiple hosts. For an extensive qualitative and quantitative comparison of different bodies of work in ROS security see [8].

We propose two different approaches for securing ROS using post-quantum secure schemes: one at the application level using the open source liboqs [2] library and the other at the network layer which is based on an integration of the Secure ROS library and a modification of IPSec using quantum secure schemes[1].

### B. Post Quantum Secure Encryption Methods

As mentioned previously, modern communication is strongly reliant on the RSA algorithm [1] for asymmetric encryption. Asymmetric ciphers have a higher computational cost when compared to standard symmetric schemes (which use the same keys for encryption and decryption of data). However, unlike symmetric ciphers, asymmetric approaches do not require an *a priori* exchange of secret keys. Under the asymmetric approach, the message for an intended recipient is encrypted using their public key. Once encrypted, the message can only be decrypted by an entity in possession of the associated private key. The standard implementation of setting up a secure communication channel involves using an asymmetric cipher to exchange keys for symmetric encryption.

The guarantee of asymmetric encryption is based on the hardness of the underlying problem used to generate the keys. For example, the RSA algorithm is based on the hardness of the integer factorization problem. The public key that is used to encrypt messages contains the product of two large primes that are uniquely generated. If one were able to compute the factorization of the product, it would be easy to generate the private key. Thus, the security of RSA relies on the lack of efficient solutions for the integer factorization problem.

The state-of-the-art classical algorithm (general number field sieve) for factoring integers is subexponential [9]. Thus, by using randomly generated large primes, one can generate secure keys. The state-of-the-art algorithm for factoring an integer on quantum devices is Shor's algorithm that has a runtime that is polynomial in $log(N)$ where N is the integer of interest [10]. Thus, the emergence of quantum computers poses a significant threat to modern implementations of secure communication. We note that symmetric ciphers such as advanced encryption standard (AES) are also vulnerable to quantum computers. However, the impact of quantum devices on these schemes is muted. In particular, Gover's algorithm provides an $O(\sqrt{N})$ improvement over classical cryptanalysis attacks on symmetric schemes. Thus, in a post-quantum world one expects to recover the prior level of security by simply doubling the key size.

The development of post-quantum secure encryption methods for asymmetric ciphers has become a very active research area. In fact, the National Institute of Standards and Technology (NIST) is presently conducting a competition on the development of post-quantum secure ciphers [11]. Potential approaches include supersingular elliptic curve, lattice, and multivariate based encryption methods. For more details, we refer the reader to [11]. Under this competition, implementations of lattice and learning with error (LWE) based ciphers are some of the most promising entries.

Lattice based cryptography [12], relies on the replacement of integer factorization by an NP-hard problem defined on lattices [13]. The basis to generate any lattice are non-unique i.e. one can either use a good "short" basis or a bad "long" basis to generate the same lattice. Given a long basis to generate a lattice, the problem of computing a short basis (within a constant factor) is NP-hard. Thus, one can define a series of problems on the lattice that are NP-hard. Examples include:

*a) Shortest vector problem (SVP): given a norm and a basis, the goal is to find the shortest vector in the lattice* L

---
[1] https://www.strongswan.org/

$$\min_{v \in L} ||v||$$

*b) GAPSVP: distinguish between cases where the norm of the shortest vector is less than 1 or greater than β*

*c) Closest vector problem (CVP): Given a vector that does not lie in L, find the closest vector in the lattice.*

Note that the above problems have γ-approximation versions. The primary reason that lattice based problems are particularly attractive for encryption is that they are average case-hard as opposed to most problems that are hard in worst-case [14]. Given the underlying hardness of these problems, they are unlikely to be solved by quantum computers in polynomial time, and are consequently expected to be quantum resistant.

Another promising approach for post-quantum secure (PQS) encryption is *ring (LWE)* [15]. In these problems, one has to infer the values of $s$ given the results of $As + \varepsilon \mod(q)$. In the absence of the noise/error term, the problem can easily be solved using Gaussian elimination. However, in the presence of the noise term, the problem can be shown to be as hard as several worst case lattice problems [15].

Given the activity in the PQS space, there are numerous software libraries that implement the various lattice and LWE based solutions for post-quantum secure encryption. In fact, multiple lattice based solutions have successfully entered the final round of the NIST competition on post-quantum secure encryption methods. We explored multiple libraries for integration with ROS. In particular, we tested, a) qTesla: a family of LWE based PQS digital signature schemes, b) BLISS: a lattice based PQS digital signature scheme, c) NTRU: a lattice based PQS cryptosystem for encrypting and decrypting information [16]. We now integrate these methods at both the application as well as the network layers and compare performance with respect to state-of-art techniques.

## III. APPLICATION LAYER SECURITY

The basic idea behind implementing security measures in the application layer of the ROS-based system is to prevent the need for changing the ROS middleware, and address security as an application dependent measure. For instance, in our case we only care about verifying the identity of the issuer of motion commands before executing them. Hence, we implement a signature scheme within our application without data encryption as it is not required to keep the content of the messages a secret in the given use-case.

The possible attack vectors on ROS applications range from unauthorized publishing and data access to denial of service (DoS) attacks on ROS nodes [4]. If an attacker gains access to the network, it can easily access the contents published on a topic by other nodes and can also publish malicious data on any topic. It can also flood topics with garbage data which can lead to nodes crashing and denial of service. In the specific scenario of mobile vehicle C&C, it is important to verify the identity of the issuer of the commands prior to their execution.

Digital signatures are a standard way of verifying the authenticity of the sender using asymmetric encryption techniques. A public key infrastructure (PKI) supports the distribution of public keys and the identity validation of individuals or entities with digital certificates via a certificate authority (CA). A CA is a trusted third party that either generates a public/private key pair on behalf of an entity or verifies the association of an existing public key to an entity. Once a CA validates someone's identity, they issue a digital certificate that is digitally signed by the CA (e.g. X.509 certificates) which can then be used to verify an entity associated with a public key when requested.

The sender signs the message using their private key and sends the message along with the signature. The receiver then retrieves the message from the signature using the sender's public key and matches it with the original messages. The authentication would fail in the event of a mismatch between the public and private keys or message tampering.

Although we assume that the public verification key of the sender already exists in the database of the receiver, a challenge-response authentication procedure such as the one described in [5] can be used to enable registration of new publishers and subscribers in the application. The application-layer security framework is depicted in Fig. 1.

### A. LIBOQS

The Open Quantum Safe (OQS) project is an open-source software project for prototyping quantum-resistant cryptography, which includes liboqs, a C library of quantum-resistant algorithms that implements several key-exchange and signature schemes [2]. This enables users to experiment with quantum-resistant cryptography. Many quantum-resistant schemes are also based on mathematical problems that are, from a cryptographic perspective, quite new, and thus have received comparably less cryptanalysis. The OQS project provides a benchmarking platform for upcoming post-quantum safe algorithms and aids cryptanalysis. Several projects have used liboqs in evaluating the performance of post-quantum schemes into new applications that currently use classical algorithms such as RSA. The work in [17] identifies two good signature candidate algorithms from NIST's Round 2 list for TLS 1.3 authentication and compares them to RSA3072.

In first integration, we use the lattice-based post-quantum secure signature scheme called qTesla, in the application layer of a ROS-based system. We compare its performance to RSA2048 in the same application.

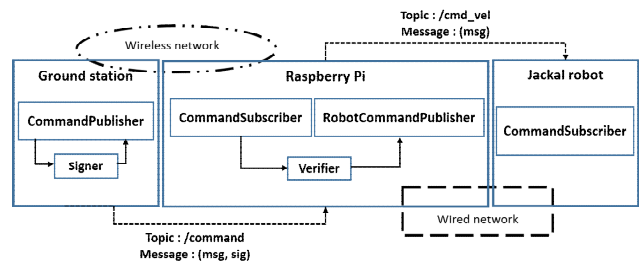

Figure 1. Framework for testing application level security

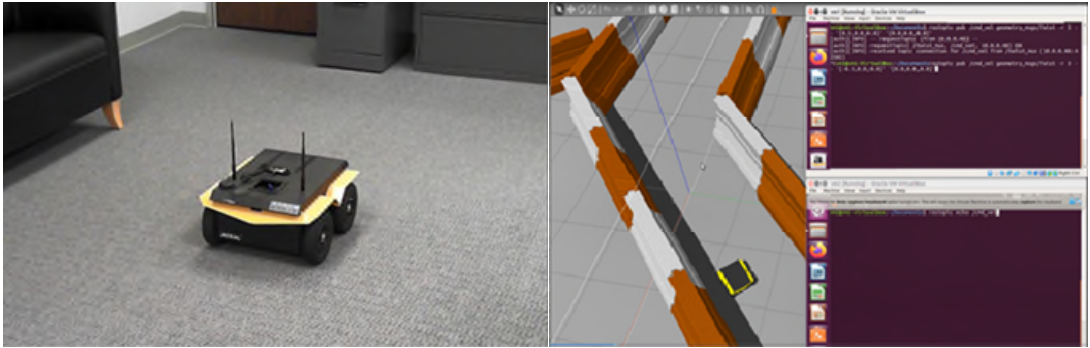

Figure 2. Mobile robot used in experiments to demonstrate post quantum secure communication in real environment (left) and in simulation (right)

*B. Demonstration*

We demonstrate our approach using a simple ROS application for C&C of a mobile robot from a ground station. The ground station in our case is a Linux machine running Ubuntu 16.04, Intel® Core™ i5-6200U CPU @ 2.30GHz × 4 with a 7.4 GiB memory. A *Clearpath Jackal AGV* (Fig. 2) is controlled remotely from the ground station and executes velocity commands. To demonstrate the feasibility of using PQS signature schemes on lightweight platforms, we use a Raspberry Pi Zero W running a 1GHz single-core ARMv6 CPU with 512MB RAM, as a relay node that performs signature verification before conveying the command to the mobile robot. If the verification fails, the message is dropped. The Raspberry Pi and the Jackal are on a separate isolated wired network, such that no malicious entity can send commands directly to the mobile robot. Fig. 1 shows the high-level application architecture for our approach.

*C. Results and comparison*

We compare the performance of our ROS integration with *qTesla* as the signature algorithm with traditional RSA2048 authentication. Fig. 3 depicts the average time to sign/verify messages using the two approaches as a function of message size. To demonstrate the performance on lightweight embedded platforms, we performed the same comparison on the Raspberry Pi Zero W and found the results to be quite satisfactory given the low computation power of the Pi (see Fig. 4).

We found that although the time to sign/verify for PQS signature is longer for larger message sizes (still in the millisecond range), it remains comparable to RSA for smaller message sizes (up to 100KB). This analysis shows the feasibility of using post-quantum authentication in real-time robotic applications. Moreover, our implementation on the lightweight Raspberry Pi Zero W is promising for applications that have restrictions on computational power and resources.

IV. NETWORK LAYER SECURITY USING SECURE-ROS

Adding security mechanisms to the application layer of ROS might not be possible if the developers have limited cyber-security knowledge. The application-level approach requires existing ROS nodes to be rewritten to incorporate security protocols. While this is not a huge burden for small applications that do not have external package dependencies, a more transparent solution might be required for complex applications.

In Section II, we discussed the body of work that deals with modifying the core communication infrastructure of ROS to incorporate security mechanisms. One such project is Secure-ROS, which was created by SRI International as a fork of ROS that enables secure communication among nodes. The operation of Secure-ROS can be broken into two distinct parts: application level authorization enforcement, and network level authentication and encryption. At the application level, the user may specify a set of authorized subscribers and publishers to topics, setters and getters to parameters, and providers (servers) and requesters (clients) of services in an authorization configuration (YAML) file for the ROS master at run time. Secure ROS will only allow authorized nodes to connect to specific topics, services, and parameters listed in the configuration file. It supports both ROSPY and ROSCPP and offers simplicity and transparency to the common user by not requiring any changes in the application itself. To further protect the ROS system against malicious attacks and ensure IP packets are not tampered or spoofed, Secure ROS relies on an Internet Protocol Security (IPSec) based Host-to-Host VPN to ensure secure authentication and encryption. IPSec is a set of open standards that ensure secure communication over Internet Protocol (IP) networks. It operates at the network layer, meaning its effectiveness is not tied to an application. After an IPSec VPN is set up, all traffic that match a set of rules defined by the user is secured before being sent over the network. This provides a significant advantage over application-level encryption since it only needs to be configured once for a given ROS system.

IPSec has two primary modes of operation, namely *tunnel* and *transport* modes, which can be configured to provide confidentiality, authentication, and data integrity checking. In tunnel mode, IPSec is used to secure traffic between two gateways, such as a router or a switch. This provides the ability for two networks to communicate securely with one another, and is the most common set up for a traditional VPN solution. In this configuration, the identity of the original sender is hidden; only the address of the gateway responsible for encryption is visible. The downside is that only outbound traffic is encrypted. This means that an IPSec VPN running in tunnel mode would not provide any means of security from malicious devices that have gained access to the internal network. The other mode of operation, transport mode, provides the ability to secure traffic being sent between hosts; however, it does not hide the identity of the sender. Secure ROS uses IPSec in transport mode. The operation of IPSec can be broken down into four basic steps. The first step is to authenticate the identity of the other device. There are multiple ways to accomplish this: pre-shared keys, digital signatures, public key encryption or external authentication. One of the more common being the use of digital signatures via RSA certificates [18]. After the two machines have been mutually authenticated, they need to agree on a set of rules that dictate which encryption algorithms can be used throughout the life of the connection. These parameters are defined in security associations (SA). For each new connection, a new security association must be negotiated and agreed upon by both parties. After the SA has been established by the two machines, they must exchange the private keys to be used for encryption. This is accomplished by the use a Key Exchange Algorithm such as IKEv2. Once the two entities have exchanged private keys, they can then begin the final step of encrypting traffic using the algorithm agreed upon in the SA.

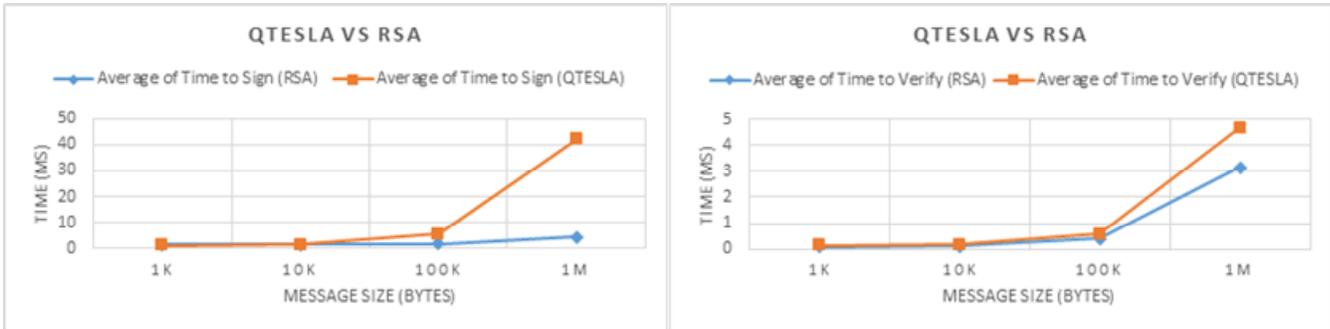

Figure 3. Comparison of time to sign and verify for different message sizes for qTesla and RSA signatures on the ground station

Secure-ROS provides a set of tools that automatically generate IPSec configuration files for each candidate machine along with its IP address specified by the user within a YAML file. These tools rely on *racoon,* the default Linux key management daemon to generate configuration files that specify the encryption algorithms in the SA, a public and private key pair unique to each host, as well as the unique public key of other hosts to be used for authentication. In its default configuration, RSA signatures are used for authentication, IKEv2 for key exchange and 3DES for data encryption. After the required files are securely transferred to each machine, communication can then be performed between the hosts specified in the YAML configuration using an IPSec VPN running in transport mode. This can be seen as access configuration at machine level, where only the hosts specified in the IPSec configuration file can get access to the ROS network.

### A. Quantum computing risks to Secure ROS

The largest threat quantum computing poses to cryptography is the ability to exhaust the key space of existing cryptographic functions within a reasonable amount of time. There are three areas within an IPSec VPN that are most vulnerable to these types of attacks:

- *The security of the private key used by each host to prove their identity via digital signatures or certificates*: Once a quantum computer is able to break the private key, it would allow a malicious node to impersonate another machine on the network, paving the way for Man-In-The-Middle attacks and allowing a malicious entity to give the mobile agent false commands [19]. Secure-ROS uses RSA signatures for authentication, which are quantum vulnerable due to Shor's algorithm.

- *The security of the shared secret key used during symmetric encryption to encrypt data being sent over the network:* Should this get compromised, an attacker could decipher the traffic being sent until the keys are renegotiated. Secure-ROS uses 3DES which has been deprecated in favor of AES as of 2018 and has been disallowed by NIST for IPSec VPNs.

- *The security of the private key used by the key exchange algorithm to securely transport the key used by the symmetric encryption algorithm*: Should the key exchange be compromised, an attacker would be able to decipher the traffic being sent until the keys are renegotiated. This bypasses the complexity of the symmetric private key, meaning the overall security of the encryption algorithm is dependent on strength of both the encryption and key exchange algorithms. Secure-ROS uses IKEv2 which relies on Diffie-Hellman key exchange with a modulus group of 1024, and can be broken by a quantum computer [20].

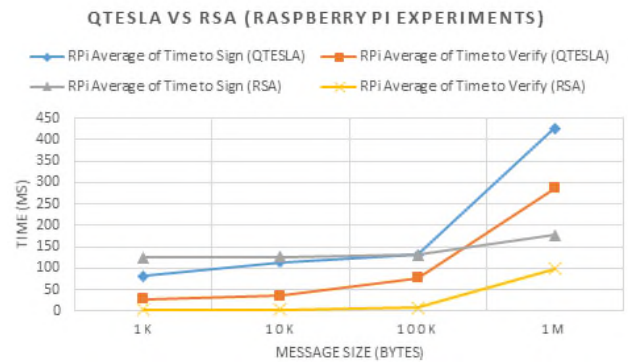

Figure 4. Comparison of time to sign and verify for different message sizes for qTesla and RSA signatures on the Raspberry Pi Zero W

Table 1. Secure ROS authorization configuration for demonstration

| Topic | Ground station | | Monitoring agent | | Attacker | |
|---|---|---|---|---|---|---|
| | Publish | Subscribe | Publish | Subscribe | Publish | Subscribe |
| /command | ✓ | ✓ | ✗ | ✗ | ✗ | ✗ |
| /e-stop | ✓ | ✓ | ✓ | ✗ | ✗ | ✗ |
| /status | ✓ | ✓ | ✓ | ✓ | ✗ | ✗ |

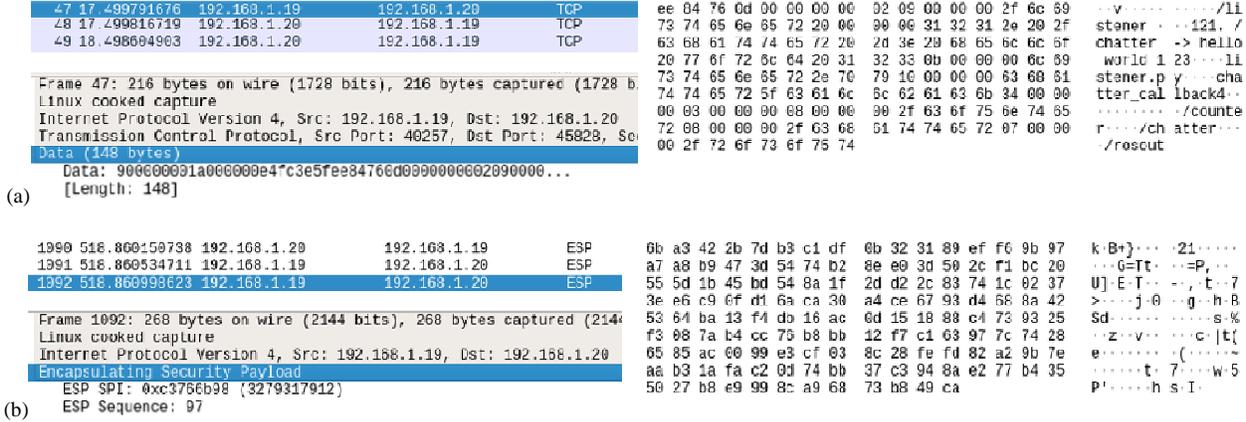

Figure 5. (a) This figure shows the Wireshark output from the attacker's perspective when there is no security mechanism present. Data in the TCP packet sent by ground station can be seen by an adversary in plaintext
(b) This figure shows the Wireshark output from the attacker's perspective once the IPSec tunnel is created. Traffic is encapsulated by ESP header and underlying TCP packet is not visible to an adversary

### B. Integrating Quantum Resistant Algorithms into Secure ROS

Due to the separation between the application level authorization enforcement and the network level authentication and encryption, it is possible to change the network-level security mechanisms of Secure-ROS without modifying core ROS packages. As *racoon* does not support any quantum resistant authentication or key exchange methods, we need to replace it with a new IPSec VPN solution called *strongSwan*[2] which supports lattice-based encryption and signature algorithms.

In order to strengthen the authentication aspect of Secure-ROS, we replaced the RSA signatures with Bimodal Lattice Signature Scheme (BLISS) certificates. BLISS is a modern cryptosystem conceptualized in 2013 that is projected to be resistant to quantum-computing attacks [21]. Akin to RSA, BLISS is an asymmetric encryption scheme used to verify the identity of a given host. BLISS certificates are handled the same way as traditional certificates, except the public and private key pair are created using lattice based schemes.

In order to implement BLISS certificates into the ROS network, a self-signed Certificate Authority (CA) must first be created on the ROS ground station. StrongSwan comes bundled with a public key infrastructure (PKI) tool that allows the user to easily generate BLISS public and private keys. The first step to creating a CA is to create the private CA key. This key can then be used to create a CA certificate which contains the public key paired with the private CA key. The certificate is then installed on each of the machines on the network. Since the public CA key present in the CA certificate can only be used to decipher data encrypted using the private CA key, entities on the network can confirm the identity of another machine by validating the signature of their certificate using the public CA key contained within the preinstalled CA certificates. If the machine is able to successfully decrypt the certificate, this indicates the certificate was signed by the same CA and is part of the trusted network.

With a quantum resistant method of authentication in place, the next point of weakness to address is the security of the symmetric encryption method. Fortunately, AES is quantum secure [22] when paired with a large enough key size (AES-256 for medium-term and AES-512 for longer term) and is strongSwan compatible.

In addition to its ability to utilize BLISS certificates, strongSwan is also compatible with NTRU for key exchange. NTRU is a lattice-based post-quantum encryption algorithm that is expected to be quantum resistant [21]. StrongSwan features an implementation based on the NTRU-Crypto C source code which acts as a replacement for the traditional Diffie-Hellman key exchange used in IKEv2. In addition to the added security, NTRU also boasts faster computation time over RSA which helps mitigate the performance impact of adding encryption.

### C. Security enhancements over Secure ROS and IPSec

After an IPSec tunnel is established, Secure ROS operates without impact on the standard work flow. To confirm the effectiveness of the IPSec VPN we use two virtual machines to communicate with each other over an unsecured network, while a third machine acts as an attacker. Without a VPN, the content of the traffic being sent between the two machines is visible to anyone on the network. To illustrate this, we use a simple ROS publisher node to periodically send a "hello world" message along with a counter displaying the number of messages sent.

By using the packet capture program, Wireshark, we analyze the traffic being sent by ROS without a working VPN. In figure 5(a) we show the ROS traffic being sent in a TCP packet. By digging into the TCP packet, one can see that the message being sent to the other ROS node is visible in plain text.

On the other hand, when the VPN is successfully configured, the traffic is encapsulated in an Encapsulated Security Payload (ESP) header. The ESP header hides the true protocol and underlying data behind a layer of encryption. This means that should an adversary intercept a message meant for a robot, they will be unable to decrypt the data being transmitted (Fig. 5b).

IPSec configuration files are used to control the connection and specify parameters negotiated in the SA, such as the mode of operation and the method of authorization, encryption and key exchange. The header contains information that is relevant to all connections and the body contains information relevant to a single connection. The user can specify which cryptographic functions are allowed to be used for different cases. In our example, we used a "trap all" method; this means all traffic, both in- and outbound, must be first secured through an IPSec tunnel. StrongSwan will attempt to establish a VPN for every new connection. Normally this method acts as a hindrance since it prevents any unsanctioned connections from outside the network, effectively blocking Internet connectivity. In our case, the robots operating on the network do not need to communicate outside the network; thus this limitation is not a drawback. In our configuration, BLISS signatures are used for authentication, NTRU-192 is used for key exchange and AES-256 is used for data encryption.

Using the IPSec configuration files, depending on the application requirements, the security mechanisms can be modified. For example, if responsiveness is of a higher priority than encryption strength, AES-512 can be changed to use a smaller key size by simply changing the parameter to AES-256 or AES-128. A weaker encryption method can still be quantum resistant if the confidential life time of the data is short. It is also possible to change the configuration file to only establish secure connections when attempting to connect to a given range of IP addresses. This would allow for the robot to maintain Internet connectivity, if needed, while still communicating with the ground station via an IPSec tunnel.

### D. Demonstration details

To demonstrate how the security infrastructure described above applies to command and control of mobile agents, we used a Gazebo simulation environment (Fig. 2) with a simulated mobile robot. The different components of the application are: a ground station, a mobile agent, a monitoring agent, and an attacker present in the network. We verified the security mechanisms at both the application layer (provided by secure ROS) and the network layer (provided by strongSwan). The authorization configuration provided to Secure ROS ensures that only the ground station can give velocity commands to the robot and the monitoring agent can issue emergency stop commands. Unlike standard ROS (no security), these rules enhance security by blocking unauthorized commands (see Table 1) even from trusted machines, limiting the attack surfaces in the event that an attacker gains access to a trusted host. Moreover, in the event that an attacker gains access to the network, any IP packets coming from their IP address will be rejected if they are not authenticated with the secure IPSec tunnel. Fig. 5 shows the system behavior with and without network layer security, respectively.

Table 3. Time taken for initial key exchange and authorization

| | |
|---|---|
| NTRU-192 + BLISS | 0.013 s |
| Diffie-Hellman Mod Group 1024 + RSA 4096 | 1.002 s |

Table 2. Performance impact of different levels of encryption with respect to message size and frequency

| Msg Size (bytes) | Target Frequency (hz) | Actual Frequency (hz) | | |
|---|---|---|---|---|
| | | No Encryption | AES-256, SHA-512 | 3DES, SHA-256 |
| 706 | | | | |
| | 5 | 5.000 | 5 | 4.999 |
| | 50 | 49.995 | 49.952 | 49.965 |
| | 500 | 499.806 | 499.101 | 499.589 |
| 1306 | | | | |
| | 5 | 5.000 | 4.999 | 5 |
| | 50 | 49.994 | 50.001 | 49.992 |
| | 500 | 491.811 | 491.469 | 490.043 |
| 6106 | | | | |
| | 5 | 4.999 | 5.001 | 5 |
| | 50 | 50.038 | 50.059 | 50.046 |
| | 500 | 160.421 | 141.854 | 129.919 |
| 12176 | | | | |
| | 5 | 5.002 | 5.004 | 5.003 |
| | 50 | 49.993 | 49.909 | 48.891 |
| | 500 | 81.771 | 72.844 | 67.558 |
| 60502 | | | | |
| | 5 | 5.012 | 5.046 | 5.014 |
| | 50 | 17.429 | 15.668 | 13.998 |
| | 500 | 17.185 | 15.634 | 13.643 |

### E. Results and comparision

We compare the performance of our Secure ROS and strongSwan integration against the default IPSec implementation currently used with Secure ROS, as well as with a system with no security. The purpose is to show the performance impact of using post-quantum schemes for network layer authentication and encryption. We use BLISS signatures along with NTRU key exchange algorithms and compare the results with RSA and IKEv2 used in IPSec. We compare the performance impact with respect to varying ROS message size and message frequency. The analysis is done on virtual machines running Ubuntu 16.04, Intel i7-7700HQ @ 2.60GHz with 1GiB memory. The comparison of average message frequency for different security schemes is shown for varying message sizes and target frequencies in Table 2. Note that this empirical data reflects the effect of using AES-256 and SHA-512 (quantum resistant) in our application as compared to 3DES and SHA-256 used in standard IPSec. The effect of post-quantum schemes on the authentication and key exchange is reflected in the connection setup time shown in Table 3. Our post-quantum implementations are orders of magnitude faster than standard IPSec, which is a promising result.

## V. FUTURE WORK

In our application layer security implementation, we focused on the authentication aspect of our C&C application with a pre-defined set of nodes. In future work, we wish to enhance the PKI management infrastructure such that new trusted nodes can be added to the ROS network at run-time. We also wish to add encryption to the application layer to ensure messages that contain sensitive information are protected. Secure ROS currently uses plaintext YAML configuration files which can be manipulated by malicious entities without being detected. We plan to alleviate this issue by using certificates. The post-quantum safe security schemes that we used in this paper are still in their active development phase and will reach standardization in the coming years. We will continue to monitor their development and integrate the most reliable and efficient algorithms in our framework, enabling security in future ROS systems.

## VI. CONCLUSIONS

Given the vulnerability of popular asymmetric ciphers to Shor's algorithm, the rapid development of quantum computing devices poses a significant near-term threat. Thus, the integration of post quantum secure methods with existing and emerging uses is of paramount importance. In this work, we have demonstrated the integration of post-quantum secure encryption schemes such as lattice and learning with errors-based encryption with the robotic operating system. We demonstrate the above by integrating various PQS encryption libraries with ROS. This enables the secure command and control between the ground station and mobile agents. In particular, we integrate these schemes in the application and network layers. The former as a means for retrofitting existing solutions and latter for the design of new ones. We find that PQS schemes integrate successfully with ROS, and provide an attractive mechanism for dealing with the impending threat of quantum computers. We also find that the performance of PQS schemes is not a limiting factor and often desirable.